\documentclass[lettersize,journal]{IEEEtran}
\usepackage{amsmath,amsfonts}
\usepackage{algorithmic}
\usepackage{algorithm}
\usepackage{array}
\usepackage[caption=false,font=normalsize,labelfont=sf,textfont=sf]{subfig}
\usepackage{textcomp}
\usepackage{stfloats}
\usepackage{url}
\usepackage{verbatim}
\usepackage{graphicx}
\usepackage{cite}
\usepackage{xcolor}
\usepackage{tikz}
\usepackage{amsmath}
\usepackage{multirow}
\usepackage{threeparttable}
\usepackage{booktabs}
\usepackage{tabularx}

\usepackage[colorlinks=true,citecolor=blue,linkcolor=blue,urlcolor=blue]{hyperref}

\definecolor{lime}{HTML}{A6CE39}
\DeclareRobustCommand{\orcidicon}{%
	\begin{tikzpicture}
	\draw[lime, fill=lime] (0,0) 
	circle [radius=0.16] 
	node[white] {{\fontfamily{qag}\selectfont \tiny ID}};
	\draw[white, fill=white] (-0.0625,0.095) 
	circle [radius=0.007];
	\end{tikzpicture}
	\hspace{-2mm}
}

\foreach \x in {A, ..., Z}{%
	\expandafter\xdef\csname orcid\x\endcsname{\noexpand\href{https://orcid.org/\csname orcidauthor\x\endcsname}{\noexpand\orcidicon}}
}

\begin{document}

% Previous Title
% \title{Estimation of Vertical Ground Reaction Forces During Gait Using The Center of the Pressed Sensors}

% New Title Suggestions
\title{%Improving 
Reliable Vertical Ground Reaction Force Estimation with Smart Insole During Walking} 
%Accuracy Through Multi-Sensor Data Fusion and Machine Learning

% \title{A Novel Approach to Vertical Ground Reaction Force Estimation Using Multi-Sensor Data Fusion and Machine Learning}

% \title{Optimizing Vertical Ground Reaction Force Estimation Accuracy Using Multi-Sensor Data Fusion and Machine Learning}

%\author{Femi Olugbon\orcidA{},~\IEEEmembership{Student Member,~IEEE,}~and~Diliang Chen\orcidB{},~\IEEEmembership{Member,~IEEE,}
        % <-this % stops a space
%\thanks{Manuscript received April xx, 20xx; revised August xx, 20xx. (\textit{Corresponding author: Diliang Chen.)}}
%\thanks{Femi Olugbon and Diliang Chen are with the Department of Electrical and Computer Engineering, University of New Hampshire, Durham, NH 03824 USA (e-mail:femi.olugbon@unh.edu; diliang.chen@unh.edu)}}

\author{Femi Olugbon, Nozhan Ghoreishi, Ming-Chun Huang, Wenyao Xu, and Diliang Chen
        % <-this % stops a space
\thanks{Manuscript received April xx, 20xx; revised August xx, 20xx. (\textit{Corresponding author: Diliang Chen.)}}
\thanks{Femi Olugbon, Nozhan Ghoreishi, and Diliang Chen are with the Department of Electrical and Computer Engineering, University of New Hampshire, Durham, NH 03824 USA (e-mail:femi.olugbon@unh.edu, nozhan.ghoreishi@unh.edu, diliang.chen@unh.edu), Ming-Chun Huang is with the Department of Data and Computational Science, Duke Kunshan University, Suzhou, China (e-mail:mh596@duke.edu), Wenyao Xu is with the Department of Computer Science \& Engineering, University at Buffalo, the State University of New York, Buffalo, NY 14260 USA (e-mail: wenyaoxu@buffalo.edu)}}

\markboth{Journal of \LaTeX\ Class Files,~Vol.~14, No.~8, August~2021}%
{Shell \MakeLowercase{\textit{et al.}}: A Sample Article Using IEEEtran.cls for IEEE Journals}

\IEEEpubid{0000--0000/00\$00.00~\copyright~2021 IEEE}
% Remember, if you use this, you must call \IEEEpubidadjcol in the second

\maketitle

\begin{abstract}

The vertical ground reaction force (vGRF) and its characteristic weight acceptance and push-off peaks measured during walking are important for gait and biomechanical analysis. Current wearable vGRF estimation methods suffer from drifting errors or low generalization performances, limiting their practical application. This paper proposes a novel method for reliably estimating vGRF and its characteristic peaks using data collected from the smart insole, including inertial measurement unit data and the newly introduced center of the pressed sensor data. These data were fused with machine learning algorithms including artificial neural networks, random forest regression, and bi-directional long-short-term memory. The proposed method outperformed the state-of-the-art methods with the root mean squared error, normalized root mean squared error, and correlation coefficient of 0.024 body weight (BW), 1.79\% BW, and 0.997 in intra-participant testing, and 0.044 BW, 3.22\% BW, and 0.991 in inter-participant testing, respectively. The difference between the reference and estimated weight acceptance and push-off peak values are 0.022 BW and 0.017 BW with a delay of 1.4\% and 1.8\% of the gait cycle for the intra-participant testing and 0.044 BW and 0.025 BW with a delay of 1.5\% and 2.3\% of the gait cycle for the inter-participant testing. The results indicate that the proposed vGRF estimation method has the potential to achieve accurate vGRF measurement during walking in free-living environments. % fusion of the center of the pressed sensor and inertial sensor data improves the accuracy and robustness of real-time vGRF estimation in real-life walking scenarios. 

\end{abstract}


\section{Introduction} \label{introduction}

\IEEEPARstart{T}{he} vertical ground reaction force (vGRF) is the dominant component of the ground reaction force during walking, which plays an important role in human biomechanical studies, providing essential data for estimating kinematics and kinetic gait parameters \cite{ancillao2018stereo, chen2017risk, chen2018risk}.

%Due to the importance of GRF in studying human movement and gait analysis, researchers have proposed various methods for measuring and estimating GRF with varying degrees of accuracy and reliability. 
%The vGRF exerted on each foot during walking is characteristically dominated by two peaks, with the rise and fall periods of each peak taking up about half of the stance phase time \cite{watkins_chapter_2009}.

The vGRF has been measured or estimated using a variety of methods in clinical and real-life scenarios. In the traditional vGRF measurement methods, in-ground force plates and force-plate integrated treadmills are commonly used to analyze human gait due to their high accuracy, reproducibility, and reliability \cite{brandes08}. However, their complexity and size limit their applications to confined laboratory and clinical settings. The limitations of traditional vGRF measurement methods restrict their applications in real-life scenarios, such as remote monitoring of patients, performance analysis of athletes, and gait analysis for activities of daily living where accurate vGRF measurements are required. Also, complex activities such as stairs and ramp ascent and descent usually require expensive setups that comprise multiple force plates, high-speed motion cameras, and complex signal processing software \cite{stacoff07,lee08dev}.

To overcome the limitations of traditional vGRF measurement systems, wearable force sensing insole systems have been explored to achieve vGRF measurement in free-living environments \cite{chen2019bring}. %The wearable pressure-sensing insole system uses flexible insole-shaped sensor arrays to track the plantar pressure distribution \cite{chen2019bring}. 
Current designs of wearable pressure sensors use different sensing methods such as force-sensitive resistors (FSR) \cite{khandakar22design}, resistive carbon polymers \cite{chen2022optimal}, and piezoresistive sensors \cite{chen2018customizable} to measure continuous pressure values under the foot's surface. However, existing wearable insole systems require frequent calibrations due to sensor drift. This leads to inconsistent readings over time \cite{saadeh2017evaluating}.

\begin{comment}
    
the performance of existing wearable insole systems is limited due to the following limitations:

\begin{enumerate}

    \item The low reliability of existing insole systems due to sensor breakage after only a few trials reduces their durability \cite{price2016validity}.

    \item Existing wearable insole systems require frequent calibrations due to sensor drift, particularly in FSRs. This leads to inconsistent readings over time, limiting their applications to short-term usages \cite{saadeh2017evaluating}.
    
\end{enumerate}
\end{comment}
%  Break the second column text from the footer.
\IEEEpubidadjcol

To address the limitations of wearable insole systems in measuring vGRF, several methods were developed to estimate vGRF indirectly from the kinematics of human body segments \cite{ancillao2018indirect}. Inertial measurement units (IMU) have become popular in gait analysis research due to the advances in the development of small and wearable sensors capable of measuring the acceleration, angular velocity, and magnetic fields \cite{tedesco2016experimental}. Previous studies have used IMU and human biomechanical models \cite{shahabpoor2017measurement} and machine learning models \cite{ancillao2018indirect} to estimate vGRF. Martinez-Pascual et al. \cite{martinez2023estimating} used five IMU sensors mounted on the foot, lower leg, thigh, hip, and the seventh cervical bone alongside machine learning algorithms in estimating vGRF. Jiang et al. \cite{jiang2020estimating} found the optimal placement of IMU sensors for estimating vGRF during overground walking using a random forest regression (RF) algorithm. However, IMU orientation and sensor drift affect the vGRF estimation performances \cite{tan2019influence}, requiring multiple calibrations and complex filter implementations. In addition, current indirect methods of vGRF estimation do not generalize well to unseen user data during testing \cite{martinez2023estimating, jiang2020estimating}. 

To address the problems of current direct and indirect vGRF estimation methods, a novel vGRF estimation method of fusing the center of the pressed sensors (CoPS) and IMU data was proposed. To avoid the influences of drifting errors of pressure sensors, all the pressure sensors on the flexible pressure sensor array were used as pressure switches to detect if the pressure sensor was pressed or not, and CoPS is calculated as the center of all the pressure sensors of the sensor array. %pressure-sensing array previously developed in \cite{chen2019bring} was used as a switch, thereby eliminating the drift associated with insole pressure sensors. 
In addition to the popularly used artificial neural networks (ANN) and random forest regression (RF), the performance of bi-directional long-short-term memory (LSTM) in the estimation of vGRF was also evaluated. 

\noindent The key contributions of this study are summarized as follows:
\begin{enumerate}

\item Proposed a novel vGRF estimation method using IMU data and newly introduced CoPS data. CoPS is derived from the pressure sensor array by treating the pressure sensors on it as pressure switches, which eliminates the limitations of pressure sensors, such as hysteresis and sensor drift; % and enhances the accuracy and reliability of the vGRF estimation.

\item In addition to popularly used machine-learning algorithms -- ANN and RF, LSTM was introduced for vGRF estimation. The LSTM model demonstrated superior performance because of its capability to capture temporal dependencies; % and generalizing across unseen data, outperforming the commonly used RF and ANN models.
%\item Implemented sensor fusion strategies integrating IMU and CoPS data in estimating vGRF, thereby improving the prediction accuracy, robustness, and reliability over the baseline IMU estimation methods.

\item Developed experiments to comprehensively evaluate the effectiveness of the proposed system in intra-participant and inter-participant tests, showing superior generalization performance of the proposed method over the existing state-of-the-art IMU-based vGRF estimation methods.
\end{enumerate}

\noindent The remainder of this article is organized as follows. Section \ref{related_work}  provides an overview of vGRF measurement and estimation techniques. Section \ref{methods} discusses the system design, data processing, and the machine-learning model architecture used in this study. Section \ref{results} describes the experiment design and evaluation results. In Section \ref{discussion}, we discussed the limitations and future work of the proposed GRF estimation techniques. Finally, the conclusion is given in Section \ref{conclusion}.

% ******************************************************* End of Introduction

\section{Related Works} \label{related_work}
This section focuses on the contributions and limitations of different vGRF measurement and estimation methods.

\subsection{Direct Measurement Methods of vGRF} \label{direct_measurement}

\subsubsection{Traditional Measurement Methods} \label{traditional_methods}
%The state-of-the-art method to measure vGRF in common activities like single-step capture uses floor-mounted instrumented force plates (IFP). IFPs can be easily integrated with other acquisition devices, such as electromyography and optical motion cameras, providing reliable datasets for integrated and multifunctional gait analysis and evaluation \cite{ancillao2018stereophotogrammetry}. IFPs 
Floor-mounted instrumented force plates have been a popular system for gait analysis because of their high accuracy, reliability, and reproducibility in laboratory and clinical settings \cite{ancillao2018stereo,brandes08, lee08dev}. However, force plates can only collect data in a single-step manner which makes it challenging to collect a large amount of data \cite{stacoff07}. % t often require optical motion capture systems for complex motion analysis such as jumping, making implementation costly and requiring complex signal processing equipment \cite{stacoff07}.
%In-ground force plates have limited applications when repeated cyclic trials are required, such as recording multiple steps in walking or running activities. To overcome this limitation, researchers have developed force plate integrated 
The instrumented treadmill with force plates integrated can address this problem by allowing the recording of multiple steps during cyclic activities \cite{van2017real}. However, instrumented treadmill is designed to be used in controlled environments, limiting their application in real-world settings.
%In addition, changes in walking patterns and strategies have been reported when walking on in-ground force plates and instrumented treadmills \cite{van2014overground,najafi2011laboratory}, affecting the user's gait patterns.

%Platform-based pressure mats were proposed to address the limitations of IFP and IT. Pressure mats use a portable matrix of pressure or forces to measure the vGRF and pressure distribution under the foot \cite{novel_measurement,zebris_mat}. Although pressure mats can provide accurate GRF and center of pressure measurements for static and dynamic activities in outdoor environments, they require flat surfaces or platforms to operate, limiting their everyday life applications. Furthermore, the foot contact area must lie in the center of the sensing area to ensure accurate readings, requiring the users to familiarize themselves with the system and potentially affecting their natural gait \cite{macwilliams2000clinical}.

%While traditional methods serve as the gold standards for measuring GRF, they are expensive to set up and require highly skilled operators. 

\subsubsection{Wearable Sensor-Based Measurement Methods} \label{wearable_methods}
Advancements in wearable technologies enable solutions for vGRF measurement in free-living environments. %are driving the development of miniature, lightweight, and energy-efficient sensors suitable for wearable systems, increasing research focus on real-time health monitoring. One area that has attracted significant contributions is the analysis of human gait using wearable pressure and force sensing systems sandwiched between the plantar surface and the shoe sole. 
Insole-based pressure sensing systems have been proposed by various studies. %to address the limitations of traditional gait analysis methods, leading to the development of multiple systems using different sensor technologies. 
%Resistive-based sensing systems are commonly used for monitoring the plantar pressure distribution in real-time \cite{chen2019bring,chen2022optimal,chen2020ubiquitous,liu2010wearable}. 
For example, pressure sensors based on piezoresistive technologies \cite{chen2019bring,chen2022optimal,chen2020ubiquitous,liu2010wearable, chen2018customizable, measurement_specialties, pcb_electronics}, capacitive sensing methods \cite{novel_measurement,mertodikromo2020low,tang2023wearable}, inductive sensors \cite{liu2009small,wang2022portable}, and optical sensors \cite{mcgeehan2021optoelectronics,tavares2021optically} have been explored to develop insole system for measuring vGRF under foot.
However, existing flexible pressure sensors used for insole development suffer from hysteresis problems and drifting errors, which limits their applications in real-world \cite{saadeh2017evaluating, razak2012foot}

\begin{comment}
    
Although insole pressure-sensing systems provide a low-cost, wearable, and low-power alternative for measuring GRF compared to traditional methods, they face several challenges:

\begin{itemize}
    \item Different loading and unload paths of pressure sensors cause hysteresis problems \cite{saadeh2017evaluating}.
    
    \item Nonlinear relationship between pressure and the sensing element requires complex signal processing circuitry, driving up the overall cost of the system \cite{saadeh2017evaluating}.
    
    \item Continuous cyclic loadings and temperature changes cause sensor drifts, leading to lower reliability and time-dependent performances \cite{razak2012foot}.
    
    \item Frequent calibration requirements due to sensor drifts limit their everyday use \cite{razak2012foot}.
    
\end{itemize}
\end{comment}

\subsection{Indirect Estimation Methods of vGRF} \label{grf_estimation_methods}

\subsubsection{Human Biomechanical Methods for vGRF Estimation} \label{biomech_methods}

Human biomechanical modeling methods have been proposed as an alternative to estimate vGRF. Optical motion capture systems and IMU-based full-body motion capture systems were used for measuring human body kinematics data \cite{karatsidis2016estimation}. These methods use three-dimensional orientation data, tri-axial linear and rotational displacements, velocity, and acceleration of each human body segment alongside a dynamic human body model to estimate vGRF \cite{shahabpoor2017measurement}. 
%Human biomechanical methods rely on the body dynamic model and inverse dynamics to estimate vGRF. 
However, these methods suffer from accumulated errors during activities such as the double support phase of walking, where a closed-loop mechanical chain is formed. This results in an impossible state for uniquely estimating the vGRF for each foot \cite{ancillao2018indirect}. 

\subsubsection{Machine Learning Approaches for vGRF Prediction} \label{ml_approach}

Several studies have proposed supervised machine learning and deep learning models to predict vGRF using different sensing technologies \cite{ancillao2018indirect, shahabpoor2017measurement}. Machine learning models, especially neural networks, can learn complex hidden patterns between input features and dependent variables \cite{almeida2002predictive}, making them suitable for vGRF prediction tasks. Predicting vGRF without explicitly formulating the human biomechanical models has been made possible by using machine learning models \cite{ancillao2018stereo}.

ANN has shown excellent performances in previous studies for gait analysis tasks \cite{lafuente1998design,wouda2018estimation}, and foot shuffling detection \cite{ansah2023smart} using the measured vGRF. Moon et al. \cite{moon2024prediction} proposed an optimized ANN model to predict vGRF using a wearable accelerometer and integrated cameras. %Zhu et al. \cite{zhu2023using} proposed a novel transformer architecture for estimating vGRF using four IMU sensors and wearable pressure sensors. %, demonstrating an improved performance over existing neural network models. 
Several studies have proposed decision tree algorithms to model vGRF prediction problems due to their robustness to outliers, nonlinear, and unbalanced data \cite{jiang2020estimating, martinez2023estimating}. Jiang et al.  \cite{jiang2020estimating} utilized a RF regressor to predict vGRF for walking activity using a shank-mounted IMU sensor. Although existing machine learning methods yielded better performances for individual subjects, they often performed poorly on unseen subject data and conditions.

\subsubsection{Foot Center of Pressure for vGRF Estimation} \label{cop_parameter}

The foot Center of Pressure (CoP) is a critical parameter in biomechanics and gait analysis, representing the point of application of the ground reaction force vector \cite{shahabpoor2017measurement}. CoP trajectories provide valuable information about balance control, postural stability, and foot function during static and dynamic activities \cite{lugade2014center}. Several studies have proposed using CoP data to estimate vGRF, aiming to overcome the limitations of laboratory-based force plate systems.
Rouhani et al. \cite{rouhani2010ambulatory} proposed using CoP data from pressure insoles to estimate complete vGRF during level-ground walking. Jung et al. \cite{jung2014ground} used CoP data pressure mats alongside joint kinematics to estimate vGRF during walking. Although methods based on CoP data showed promising results in controlled activities, they face challenges in real-world applications due to sensor drift, frequent re-calibration, and poor generalization performances across different users \cite{razak2012foot}.

To address the limitations of existing methods for estimating vGRF, a novel method was proposed by fusing CoPS and IMU data. Three machine learning algorithms -- ANN, RF regression models, and LSTM -- were evaluated for their performance in vGRF estimation in both intra-participant and inter-participant testing scenarios. %The proposed method potentially enables more robust and improved performance of vGRF estimation in real-life settings by leveraging the strengths of both the CoPS and IMU sensors.

%To address the limitations of existing methods for estimating vGRF and the need for accurate, reliable, and drift-resistant vGRF measurements across various real-world conditions, a novel vGRF estimation method utilizing the fusion of CoPS and IMU data was proposed. To fill the gap that limits the application of existing vGRF estimation methods, three machine learning algorithms -- artificial neural networks, random forest regression models, and bi-directional long-short-term memory networks -- were evaluated for personalized and generalized performances. The proposed method potentially enables more robust and improved performance of vGRF estimation in real-life settings by leveraging the strengths of both the CoPS and IMU sensors.

% ******************************************************* Section End

% %  Break the second column text from the footer.
% \IEEEpubidadjcol

\section{Materials and Methods} \label{methods}

This section specifies details of the smart insole system, the method to extract CoPS, and the algorithms for estimating vGRF.

\begin{figure}[htbp!]
\centering
\includegraphics[width=\columnwidth]{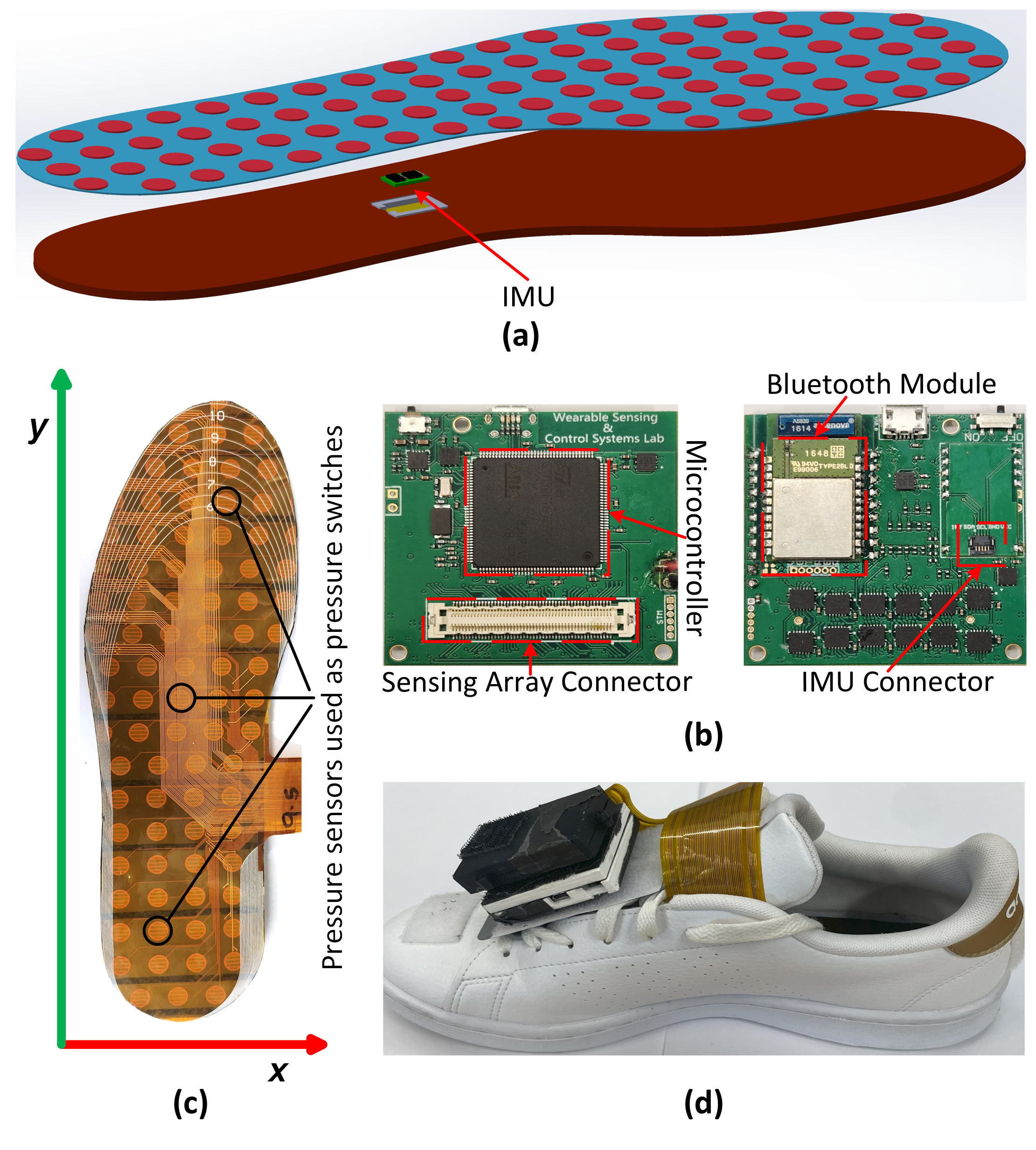}
\caption{Overview of the hardware elements of the smart insole system. (a) 3D model of the smart insole with an insole-shaped pressure sensor array on the top, an IMU in the middle, and a flexible substrate on the bottom; (b) The printed circuit board integrates the microcontroller, Bluetooth module, and the connector to the IMU and pressure sensor array; (c) The pressure-sensing array with pressure sensors uniformly distributed on it; (d) Assembled shoes with the smart insole system.}
\label{fig:system_setup}
\end{figure}

\subsection{Smart Insole System} \label{smart_insole}
%The smart insole system is important for measuring the center of the pressed sensor used in this study. 
Fig. \ref{fig:system_setup} shows the smart insole's hardware systems for signal acquisition and wireless data transmission. As shown in Fig. \ref{fig:system_setup}({\color{blue}c}), an insole-shaped pressure sensing array, is used for measuring the plantar pressure signal \cite{chen2018customizable}. %The sensing array of the smart insole system used a commercially available piezo-resistive polymer material made by EonTex to design the pressure sensors \cite{chen2018customizable}. 
The sensing array utilized a customizable design to reduce the manufacturing cost, which can be trimmed to fit foot sizes from 5.5 U.S. to 14 U.S. Fig. \ref{fig:system_setup}({\color{blue}a)} shows the structure of the smart insole, with the pressure sensor array on the top, an IMU module in the middle, and a 3D printed flexible substrate on the bottom. %methods to assemble the smart insole. The smart insole has a three-layer structure. The insole-shaped package on the first layer is the main structure of the system -- made from a flexible printed circuit board \ref{fig:system_setup}({\color{blue}a)}. The IMU sensor, placed at 40\% of the foot length, forms the second layer, and it is protected with a 3D-printed enclosure. 
%A 9-degree-of-freedom IMU sensor (ICM20948, TDK InvenSense\texttrademark) was used to design the smart insole system. 
The IMU placed at the midfoot region was used to estimate the foot's three-dimensional acceleration, angular velocity, and angle. %Data from the IMU and pressure sensing array were acquired at a sample rate of 50 Hz. %A flexible insole-shaped substrate made from thermoplastic polyurethane material shown in the third layer serves as the housing for the pressure-sensing array and the IMU sensor. 
The smart insole system is powered by a 3.7 V, 1000 mAh battery that is mounted on the shoe using a custom 3D-printed housing attachment (Fig. \ref{fig:system_setup}({\color{blue}d)}).

% Figure \ref{fig:experiment_setup} shows the experiment setup and the equipment used during the experimental sessions. One pair of pressure sensing arrays previously developed by Chen et al. in \cite{chen2018customizable}, operated as foot contact switches, were used to acquire the center of the pressed sensor data using the method outlined in \ref{cops_estimation}. 

\subsection{Center of the Pressed Sensor Measurement} \label{cops_estimation}

Fig. \ref{fig:system_setup}({\color{blue}c})) shows the pressure sensing array made from a flexible printed circuit board. The sensing array comprises 96 individual sensors uniformly distributed on the foot plantar surface. Calculating the CoPS requires using the pressure sensors as pressure switches to detect if an individual pressure sensor is pressed. Based on our previous study, an adaptive threshold was used to accurately determine the sensor's ON and OFF states \cite{chen2019bring, chen2022optimal}. %AT was chosen due to its excellent performance in distinguishing between the swing and stance phases of the gait cycle \cite{chen2022optimal}. In addition, AT has been shown to outperform fixed threshold in determining the swing and stance phases during a walking gait \cite{chen2019bring, chen2022optimal}. 
The adaptive threshold can be estimated using equation (\ref{eqn:adaptive_thresold}).

\begin{equation}
\label{eqn:adaptive_thresold}
{\mathrm{AT_{I}}} = \frac {1}{n_{\mathrm{ swing}}}\sum _{k = 1}^{n_{\mathrm{ swing}}} {\rm Pressure_{k_{I}}} + 3*\sigma _{\mathrm{ swing}}.
\end{equation}

\noindent Where $AT_{I}$ indicates the adaptive threshold of sensor $I$ (I = 1, 2, 3, \dots, 96), $Pressure_{k_{I}}$ is the pressure measured by the individual sensor $I$ in the sensing array, $n_{ swing}$ is the total number of swing pressure samples of sensor $I$ collected when the foot is off the ground, and $\sigma_{ swing}$ is the standard deviation of the swing pressure samples. % Since the pressure sensors are operated as foot contact switches, the adaptive threshold for each sensor is used to determine whether the sensors are activated or not. 
Equation (\ref{eqn:states}) shows the estimation of the sensor states:

\begin{equation}
\label{eqn:states}
\mathrm{S_I} = 
\begin{cases}
1, & \mathrm{Pressure_I} \geq \text{$\mathrm{AT_{I}}$} \\
0, & \text{otherwise}
\end{cases}
\end{equation}

\noindent where $S_{I}$ indicates the sensor state of sensor $I$, and $Pressure_{I}$ indicates the pressure measured by sensor $I$ in the sensing array. CoPS was measured by using the sensor states and their respective coordinates, as shown in equation (\ref{eqn:cops}).

\begin{equation}
\label{eqn:cops}
\left\{
\begin{aligned}
    \mathrm{CoPS_x} &= \mathrm{\frac{\sum_{j=1}^{I} x_j \cdot s_j}{\sum_{j=1}^{I} s_j}} \\
    \mathrm{CoPS_y} &= \mathrm{\frac{\sum_{j=1}^{I} y_j \cdot s_j}{\sum_{j=1}^{I} s_j}}
\end{aligned}
\right.
\end{equation}
 
\noindent where $CoPS_{x}$ and $CoPS_{y}$ indicate the anterior-posterior and medial-lateral $CoPS$, respectively, $x_{j}$ and $y_{j}$ are the coordinates of the sensors $j$, $s_{j}$ is the state of the sensor $j$, and $I$ is the total number of the sensors.

\subsection{Data Processing} \label{data_acquisition}

\begin{figure*}[!tb]
\centering
\includegraphics[width=\textwidth]{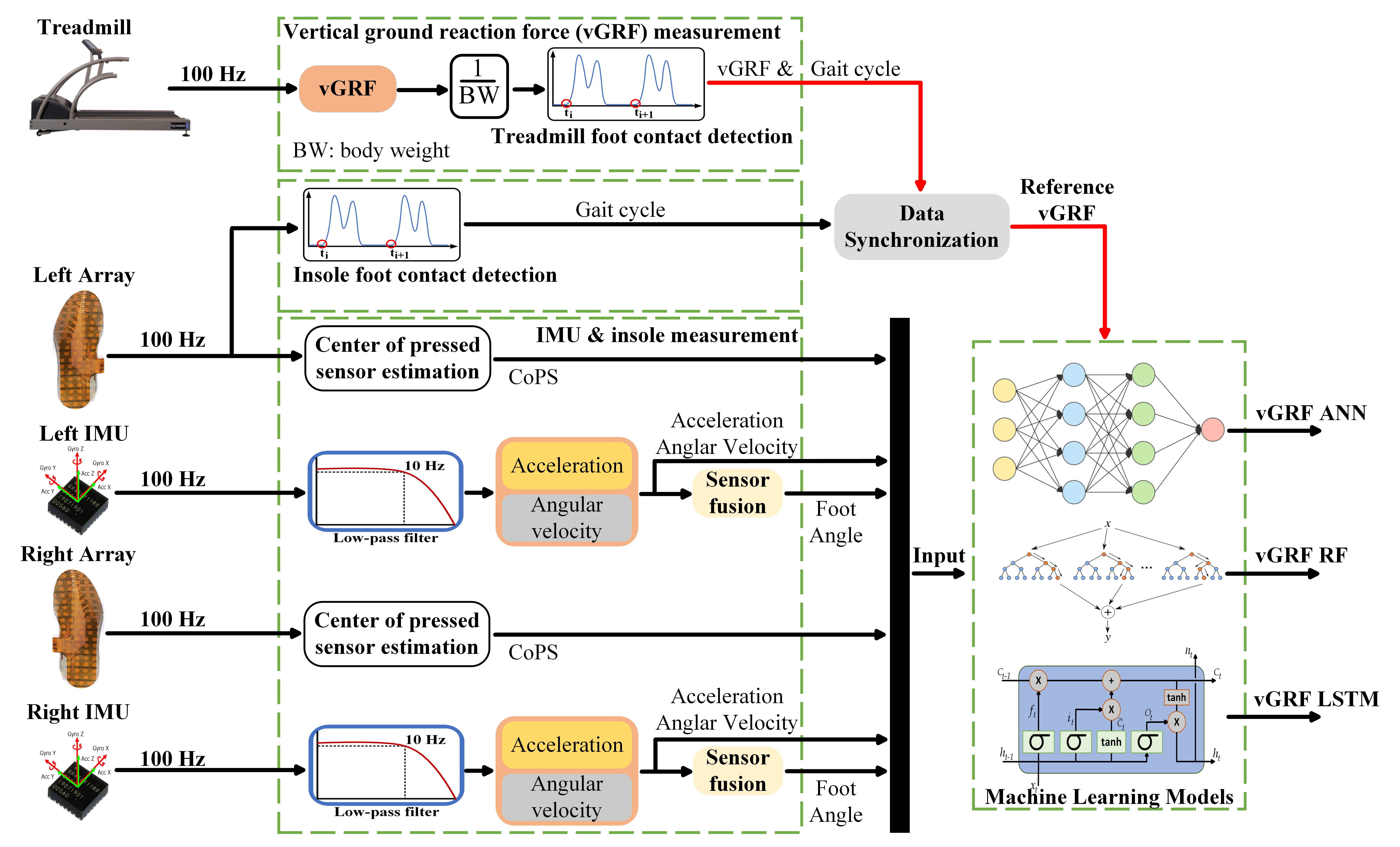}
\caption{An overview of the process for data collection, data processing, vGRF estimation.}
\label{fig:data_processing}
\end{figure*}

%and Signals are obtained from the instrumented treadmill, pressure-sensing array, and IMUs and their respective data processing before feeding them to train the machine learning models. From the treadmill, we acquired the vGRF, and it was normalized using the participants' body weight. The vGRF serves as the ground truth for the machine learning regression models. From the pressure-sensing array, the acquired pressure from individual sensors was used to estimate the CoPS as described in \ref{cops_estimation}. We measured the linear acceleration and angular velocity from the IMUs integrated into the shoe insole. The foot angles are derived from the combination of linear acceleration and angular velocity using the method described by Madgwick in \cite{madgwick2010efficient}. Gait cycle segmentation and foot contact detection were performed using the start of the stance phase (rising edge of the vGRF). Data synchronization was performed using the start of the gait cycle for the vGRF, CoPS, and IMU data. The CoPS, linear acceleration, angular velocity, and foot angles from both feet served as the input to the three machine learning models. In contrast, the normalized vGRF served as the reference input.

Fig. \ref{fig:data_processing} illustrates the detailed data processing pipeline and the inputs to the machine learning models used in this study. First, the vGRF data measured by the instrumented treadmill underwent normalization using the participants' body weight (BW). The normalized vGRF serves as the ground truth for the machine learning regression models.
The foot contact event, which can be accurately detected by both the instrumented treadmill and the smart insole, was used to synchronize the data from both systems. %Since the IMU system and the pressure-sensing array are automatically synchronized by design; their data are collected using the data acquisition hardware. 
The pressure sensors on the pressure sensor arrays were used as pressure switches, and the CoPS can be estimated with method shown in section \ref{cops_estimation}. To obtain the foot angles, we utilized the orientation filter developed in \cite{madgwick2010efficient}, using the data acquired from the linear acceleration and the angular velocities of the IMU sensors integrated into the insoles. %Once the foot angles were computed, the linear acceleration, angular velocity, and foot angle data underwent a median filter of a three-sample window and then filtered using a zero-phase low-pass filter \cite{gustafsson1996determining} at a cut-off frequency of 10 Hz.

%The start and end of each gait cycle were determined by detecting the heel-contact and toe-off events from the vGRF and pressure-sensing array systems. 
The vGRF, CoPS, and IMU data of each gait cycle (determined by two consecutive heel contact events) was padded into a window of two seconds \cite{ansah2022wearable}. All the padded data were added to the swing phase of each gait cycle. The two-second window was used because previous studies have shown that the two-second window size provides the best trade-off between speed and accuracy \cite{banos2014window}. All the processed data were further normalized to a uniform range between 0 and 1 before feeding them as inputs to the machine learning models for training and testing.

The estimated vGRF from the ANN and RF models were filtered using a zero-phase low-pass filter at a 10 Hz frequency, as the majority of the spectral component of ground reaction forces is contained below 10 Hz \cite{blackburn2016comparison}. The weight acceptance peak (WAP) and the push-off peak (POP) were detected by searching the local maxima in each stance cycle's first and second halves. The WAP determines the local maximum in the first half of the stance cycle, while the local maximum in the second half of the stance cycle corresponds to the POP \cite{chen2019bring}.

\subsection{Machine Learning Algorithms for vGRF Estimation} \label{ml_algorithms}

The vGRF estimation is a regression task. In this study, three machine-learning regression model architectures were trained to estimate the vGRF. We chose RF regression model and the ANN models due to their popularity in estimating vGRF and analyzing gait data \cite{jiang2020estimating, martinez2023estimating}. Additionally, bi-directional LSTM networks were chosen to predict vGRF due to their ability to capture temporal dependencies in both forward and backward directions. This bidirectional approach allows the model to leverage both past and future context at each time step, making it particularly suitable for analyzing the complex, cyclic nature of gait dynamics. LSTMs' capacity to model long-term dependencies and their robustness to vanishing gradient problems further enhance their effectiveness in predicting time-dependent force patterns.

The ANN, RF, and LSTM models were trained using the PyTorch Python Library (PyTorch Foundation). We used an ANN model with one input layer with the number of neurons matching the input feature size, five hidden layers with 20 neurons, and an output layer with one neuron. The RF training was performed using 200 decision trees and two leaves. The output of the RF model was majority-voted from the results produced by the individual trees. For the LSTM model, a four-layer bidirectional LSTM network with 128 units with dropout regularization was utilized to reduce overfitting. For all three models (ANN, RF, and LSTM), the RMSE loss function was used.

\subsection{Model Training and Testing} \label{ml_preprocessing}
For the training of the selected machine learning algorithms (ANN, RF, and LSTM), we assessed the model performances using two model training topologies -- intra-participant and inter-participant training methods. Intra-participant training and testing, also called same-participant testing, involves using datasets from the same user to train and evaluate the model performance. In this study, 70\% of the data from all three walking speeds of a participant is used for training, while the rest is used to evaluate the model performance on unseen data from the same user. The intra-participant testing is repeated for all participants' datasets using all three walking speeds. The model accuracy is then calculated and averaged for each walking speed across all participants.

In contrast, inter-participant testing, also known as leave-one-subject-out cross-validation, involves using all the data from a group of participants to train the machine learning model, while another dataset from a group that was not part of the training data is used to evaluate the model performance. In this study, data from seven subjects is used for training, while the data from the remaining one subject is used to test the model. This process is repeated until data from each participant has been used to test the model performance. The model accuracy is then calculated and averaged for each walking speed across all participants. The inter-participant testing evaluates the generalization capabilities of the model to an unseen population. %Intra-participant testing usually achieves better accuracy than inter-participant testing, but it always requires training new models for every new user in real-life applications.

To analyze the effectiveness of the newly introduced CoPS data and the fusion of CoPS and IMU data on the estimated vGRF, We developed three model training sets for each machine learning algorithm.

\begin{itemize}
    \item Training 1 (T1): IMU features -- three-axis linear acceleration, three-axis angular velocity, and three-axis foot angle data for both feet were included as inputs to the machine-learning models.
    
    \item Training 2 (T2): CoPS features -- anterior-posterior and medial-lateral CoPS data for both feet were included as inputs to the machine-learning models.
    
    \item Training 3 (T3): IMU and CoPS features -- data from both the IMU and CoPS from both feet were combined as the inputs to the model.
    
\end{itemize}

\subsection{Model Performance Evaluation} \label{ml_evaluation}

 %The performance of the vGRF estimation methods was thoroughly evaluated using two testing topologies: intra-participant and inter-participant testing methods. 

 This study utilized multiple performance metrics, and the performance achieved by the vGRF estimation methods was evaluated as follows:
 \begin{itemize}
     \item \textbf{Root mean squared error (RMSE)}: The RMSE measures the distance between the estimated and reference vGRF trajectories. We selected the RMSE because it penalizes larger errors, resulting in better vGRF peak estimation.
     \item \textbf{Normalized Root mean squared error (NRMSE)}: The NRMS normalizes the RMSE by dividing the RMSE by the range of the reference vGRF: equation (\ref{eqn:nrmse}), as defined by Mentaschi et al. in \cite{mentaschi2013nrmse}. A lower NRMSE in the reconstruction of vGRF indicates better model performance.
     
     \begin{equation}
     \label{eqn:nrmse}
     \mathrm{NRMSE} = \mathrm{\frac{RMSE}{max(vGRF_{ref}) - min(vGRF_{ref})}}
     \end{equation}
     
     \item \textbf{Correlation coefficient (R)}: R in this study measured the similarity estimated vGRF and the reference vGRF measured by the treadmill. The values of $\rho$ range from -1.0 to 1.0, where R = 1.0 represents a perfectly positive correlation, R = 0.0 implies no linear relationship between the reference measurement and the estimated vGRF, and R = -1.0 results in a perfectly negative correlation. In this study, Pearson's correlation coefficient was used as a metric, and the closer R gets to 1.0, the better.
     \item \textbf{WAP and POP error}: A lower RMSE between the reference and estimated vGRF peaks (WAP and POP) results in a better model performance. WAP and POP are the local maxima in the first and second halves of each vGRF stance phase, respectively \cite{chen2019bring}. %to determine the WAP and POP restively. We defined the start and end of a stance phase using the adaptive threshold outlined in (\ref{eqn:adaptive_thresold}).
     \item \textbf{WAP and POP delay}: The timing occurrence of the detected WAP and POP were also recorded, and their respective delay or advance errors were estimated. A delay or advance close to 0\% of the gait cycle implies a better performance of the model.
 \end{itemize}

%\subsection{Statistical Analysis} \label{stat_analysis}
%We performed statistical data analysis to compare the performances of all the trained models in estimating the vGRF using the validation and test data. At first, we compared the nine trained models (ANN-IMU, ANN-CoPS, ANN-Fusion, RF-IMU, RF-CoPS, RF-Fusion, LSTM-IMU, LSTM-CoPS, and LSTM-Fusion) using their RMSE and NRMSE as the evaluation metrics. We selected the best models (LSTM-CoPS and LSTM-Fusion) to perform further statistical analysis, considering the evaluation metric outlined previously. For each model, we compared the vGRF estimation accuracy, peak reproduction accuracy, peak advance or delay error, and the slope of the vGRF using the test data for both the intra- and inter-participant testing methods.

%We performed a two-way analysis of variance (ANOVA) to examine the effect of the independent variables -- sensor data type (IMU data, CoPS data, and fusion of IMU and CoPS data) and machine-learning algorithm type (ANN, RF, and LSTM) on the dependent variables -- RMSE and correlation coefficient (R), for both the intra-participant and inter-participant testing methods. Post hoc pairwise comparisons (Turkey's honestly significant difference) were further used to analyze the significant levels of the independent variables on the dependent variables using a significance level of $\mathrm{p < 0.05}$.

% ******************************************************* Section End

\section{Experiments and Results} \label{results}

\subsection{Experimental Setup} \label{experiment_setup}

\begin{figure}[!thb]
\centering
\includegraphics[width=\columnwidth]{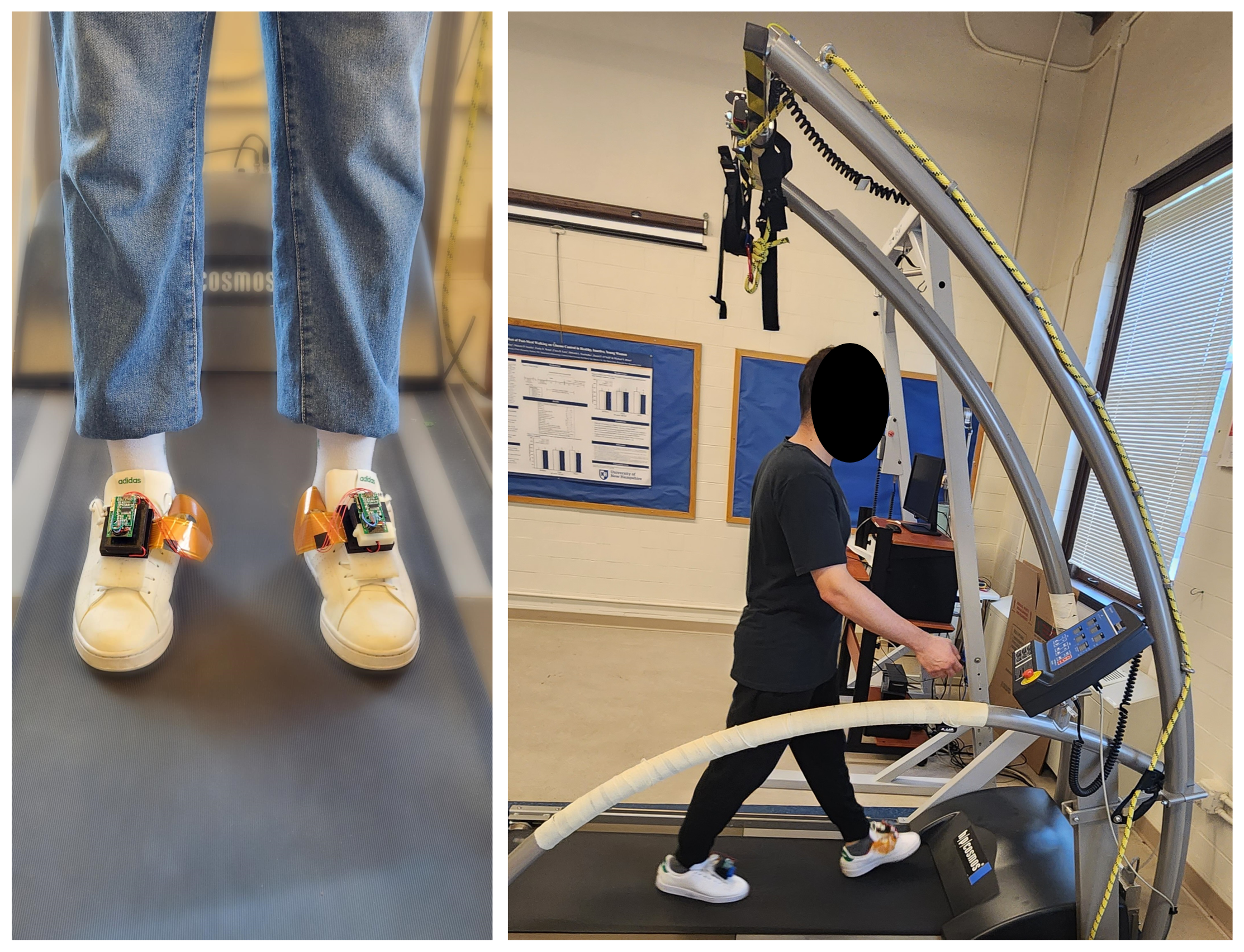}
\caption{The experiment setup. The participant wears a pair of smart insole and walks on a force-plate instrumented treadmill.}
\label{fig:experiment_setup}
\end{figure}

% \begin{figure}[tbp!]
%     \centering
%     \includegraphics[width=\columnwidth]{files/data_processing_plot.png}
%     \caption{Sample of the collected raw IMU and CoPS signals for a segment of three gait cycles, alongside the reference vertical ground reaction force (vGRF).}
%     \label{fig:raw_data}
% \end{figure}

Fig. \ref{fig:experiment_setup} shows the experiment setup. An instrumented treadmill with embedded force plates (h/p/cosmos GaitWay III) was used to measure the reference vGRF with a sample rate of 100 Hz during the experiment. 
Eight healthy adults recruited from the local university population participated in this study. The heights range between 1.66 m to 1.82 m (173 $\pm$ 5.6 cm), with foot lengths ranging from 25 cm to 27.3 cm (26.19 $\pm$ 0.65 cm) and body weights between 66.5 kg and 82 kg (75.69 $\pm$ 6.24 kg). IRB approval and written informed consent from all the participants were obtained for this study. 
During the experiment, each participant wore a pair of smart insole and walked on the treadmill at three speeds: 0.7 m/s, 1.0 m/s, and 1.4 m/s. The speeds cover the range of average gait speeds for adult humans \cite{weber2016differences}. Each trial lasted for 90 seconds, progressing from low to high speeds with a two-minute rest between each trial. Before data collection, participants were given time to warm up and become comfortable on the treadmill. The pressure sensor array and inertial measurement unit integrated into the smart insole were recorded at a 50 Hz sampling rate and upsampled to 100 Hz to match the treadmill. %Data collection began 30 seconds after the treadmill reached each set speed, allowing the treadmill's initial acceleration phase to complete.
%single-limb and double-limb support activities, each lasting for 10 seconds, to obtain the calibration data for estimating the body weight (BW) and the adaptive swing threshold.
%After recording the calibration data, participants 

All eight participants completed the entire data collection protocol. %Each of the three walking speed trials per participant lasted for 90 seconds, and data were collected at a 1000 Hz sampling rate from the treadmill, IMU, and CoPS. 
A total of 27,000 IMU and CoPS samples were collected from each participant across 3 experiment trials, resulting in 216,000 total samples collected from all eight participants. We excluded some gait cycles from the data due to incorrect/missing recordings (e.g., on some occasions when the participants walked on a single force plate embedded with the treadmill). 

% Fig. \ref{fig:raw_data} shows a sample of the data collected from the IMU and CoPS signals and their corresponding reference vGRF for three gait cycles.

\begin{table*}[htb!]
\centering
\begin{threeparttable}
\caption{Mean and standard deviation of intra-participant testing RMSE for the ANN, RF, and LSTM models. }

\begin{tabularx}{\textwidth}{|l|c|>{\centering\arraybackslash}X|>{\centering\arraybackslash}X|>{\centering\arraybackslash}X|>{\centering\arraybackslash}X|}
\hline

\multirow{2}{*}{\textbf{Model}} & \multirow{2}{*}{\textbf{Features}} & \multicolumn{4}{c|}{\textbf{RMSE (BW)}} \\ \cline{3-6}

  &  & \textbf{Global} & \textbf{0.7 m/s} & \textbf{1.0 m/s} & \textbf{1.4 m/s}  \\  \hline

\multirow{3}{*}{ANN}  & T1  & $\mathrm{0.058 \pm 0.014}$ & $\mathrm{0.054 \pm 0.014}$ & $\mathrm{0.057 \pm 0.021}$ & $\mathrm{0.063 \pm 0.015}$  \\

                      & T2 & $\mathrm{0.064 \pm 0.015}$ & $\mathrm{0.062 \pm 0.015}$ & $\mathrm{0.054 \pm 0.020}$ & $\mathrm{0.062 \pm 0.016}$  \\

                      & T3 & $\mathrm{0.056 \pm 0.014}$ & $\mathrm{0.049 \pm 0.013}$ & $\mathrm{0.054 \pm 0.020}$ & $\mathrm{0.062 \pm 0.016}$ \\ \hline  

\multirow{3}{*}{RF}   & T1 & $\mathrm{0.033 \pm 0.004}$ & $\mathrm{0.032 \pm 0.006}$ & $\mathrm{0.032 \pm 0.007}$ & $\mathrm{0.034 \pm 0.07}$  \\

                      & T2 & $\mathrm{0.037 \pm 0.005}$ & $\mathrm{0.038  \pm 0.008}$ & $\mathrm{0.034 \pm 0.010}$ & $\mathrm{0.041 \pm 0.005}$  \\

                      & T3 & $\mathrm{0.029 \pm 0.003}$ & $\mathrm{0.025 \pm 0.004}$ & $\mathrm{0.028 \pm 0.008}$ & $\mathrm{0.033 \pm 0.006}$ \\ \hline                      

\multirow{3}{*}{LSTM} & T1 & $\mathrm{0.029 \pm 0.005}$ & $\mathrm{0.028 \pm 0.006}$ & $\mathrm{0.027 \pm 0.007}$ & $\mathrm{0.029 \pm 0.004}$  \\

                      & T2 & $\mathrm{0.027 \pm 0.006}$ & $\mathrm{0.028  \pm 0.007}$ & $\mathrm{0.025 \pm 0.007}$ & $\mathrm{0.027 \pm 0.006}$  \\

                      & T3 & $\mathrm{0.024 \pm 0.003}$ & $\mathrm{0.023 \pm 0.006}$ & $\mathrm{0.023 \pm 0.003}$ & $\mathrm{0.026 \pm 0.004}$ \\ \hline
\end{tabularx}
\begin{tablenotes}
\item Note: T1 corresponds to IMU features only, T2 to CoPS features only, and T3 to data fusion of IMU and CoPS features. Global RMSE is the average of all participants' RMSE across all gait speeds.

% $^{-a}$, $^{-b}$, $^{-c}$, and $^{-d}$: items labeled $^{-a}$ are significantly different from items labeled $^{-b}$, $^{-c}$ or $^{-d}$ in the same column, but there is no significant difference within items in each of $^{-a}$, $^{-b}$, $^{-c}$, and $^{-d}$.
\end{tablenotes}
\label{tab:result_intra}
\end{threeparttable}
\end{table*}

\begin{table*}[htb!]
\centering
\begin{threeparttable}
\caption{Mean and standard deviation of inter-participant testing RMSE for the ANN, RF, and LSTM models.}

% \begin{tabular}{|l|c|c|c|c|c|}
\begin{tabularx}{\textwidth}{|l|c|>{\centering\arraybackslash}X|>{\centering\arraybackslash}X|>{\centering\arraybackslash}X|>{\centering\arraybackslash}X|}
\hline

\multirow{2}{*}{\textbf{Model}} & \multirow{2}{*}{\textbf{Features}} & \multicolumn{4}{c|}{\textbf{RMSE (BW)}} \\ \cline{3-6}

  &  & \textbf{Global} & \textbf{0.7 m/s} & \textbf{1.0 m/s} & \textbf{1.4 m/s}  \\  \hline

\multirow{3}{*}{ANN}  & T1  & $\mathrm{0.088 \pm 0.017}$ & $\mathrm{0.076 \pm 0.017}$ & $\mathrm{0.082 \pm 0.031}$ & $\mathrm{0.099 \pm 0.014}$  \\

                      & T2 & $\mathrm{0.089 \pm 0.015}$ & $\mathrm{0.082  \pm 0.019}$ & $\mathrm{0.076 \pm 0.025}$ & $\mathrm{0.099 \pm 0.009}$  \\

                      & T3 & $\mathrm{0.082 \pm 0.017}$ & $\mathrm{0.069 \pm 0.012}$ & $\mathrm{0.077 \pm 0.027}$ & $\mathrm{0.092 \pm 0.018}$ \\ \hline                      

\multirow{3}{*}{RF}   & T1 & $\mathrm{0.076 \pm 0.021}$ & $\mathrm{0.073 \pm 0.029}$ & $\mathrm{0.074 \pm 0.023}$ & $\mathrm{0.077 \pm 0.023}$  \\

                      & T2 & $\mathrm{0.077 \pm 0.020}$ & $\mathrm{0.080  \pm 0.029}$ & $\mathrm{0.068 \pm 0.027}$ & $\mathrm{0.081 \pm 0.019}$  \\

                      & T3 & $\mathrm{0.070 \pm 0.019}$ & $\mathrm{0.062 \pm 0.018}$ & $\mathrm{0.067 \pm 0.025}$ & $\mathrm{0.078 \pm 0.025}$ \\ \hline

\multirow{3}{*}{LSTM} & T1 & $\mathrm{0.050 \pm 0.005}$ & $\mathrm{0.043 \pm 0.011}$ & $\mathrm{0.047 \pm 0.011}$ & $\mathrm{0.052 \pm 0.007}$  \\

                      & T2 & $\mathrm{0.047 \pm 0.008}$ & $\mathrm{0.040  \pm 0.008}$ & $\mathrm{0.040 \pm 0.014}$ & $\mathrm{0.050 \pm 0.013}$  \\

                      & T3 & $\mathrm{0.044 \pm 0.005}$ & $\mathrm{0.038 \pm 0.009}$ & $\mathrm{0.043 \pm 0.010}$ & $\mathrm{0.044 \pm 0.006}$ \\ \hline
\end{tabularx}
\begin{tablenotes}
\item Note: T1 corresponds to IMU features only, T2 to CoPS features only, and T3 to data fusion of IMU and CoPS features. Global RMSE is the average of all participants' RMSE across all gait speeds.

% $^{-a}$, $^{-b}$, and $^{-c}$: items labeled $^{-a}$ are significantly different from items labeled $^{-b}$ or $^{-b}$ in the same column, but there is no significant difference within items in each of $^{-a}$, $^{-b}$, and $^{-c}$.
\end{tablenotes}
\label{tab:result_inter}
\end{threeparttable}
\end{table*}

\subsection{Results} \label{validation_test_result}

After training all three machine learning models (ANN, RF, and LSTM) for all three training sets (T1, T2, and T3), we evaluated the accuracy of the models with the test data for both the intra-participant and the inter-participant training methods. The RMSE, NRMSE, R, WAP, and POP errors were calculated for each gait cycle. The mean and standard deviations of the RMSE for each machine-learning algorithm are shown in Table \ref{tab:result_intra}  and Table \ref{tab:result_inter} for the intra-participant and inter-participant testing, respectively.

% The mean and standard deviations of the RMSE are shown in Fig. \ref{fig:grf_error_bar}. Additionally, Table \ref{tab:result_intra}  and Table \ref{tab:result_inter} show the respective test results for the three machine learning algorithms for both the intra-participant and inter-participant training methods.

% \begin{figure*}[htb!]
%     \centering
%     \includegraphics[width=\textwidth]{files/result_rmse_graph.png}
%     \caption{Mean RMSE (BW) and standard deviation for each machine learning algorithm and training set. The graph shows the average (global) RMSE across all gait speeds and the RMSE for gait speed.}
%     \label{fig:grf_error_bar}
% \end{figure*}

The two-way ANOVA obtained for the intra-participant testing showed that there were statistically significant effects of the machine-learning algorithm type and the input feature type on the RMSE (algorithm, F(22) = 180.20, $\mathrm{p < 0.001}$; feature, F(22) = 7.56, $\mathrm{p = 0.0041}$) for all gait speeds. The ANOVA results showed that the choice of algorithms and the types of input features used to train the models individually affect the accuracy and reliability of the vGRF estimation. %The ANOVA results showed no significant effect of the interaction between the algorithm type and feature type on the RMSE and $\mathrm{p < 0.001}$. The lack of significant effects of the interaction between the algorithm type and feature type suggests that the contribution of each factor to the variance observed in the vGRF RMSE is not dependent on each other, showing the robustness of the selected features across different machine-learning algorithms.
In the pairwise comparisons, significant differences were observed between ANN and RF ($\mathrm{p < 0.0001}$), ANN and LSTM ($\mathrm{p < 0.0001}$), and RF and LSTM ($\mathrm{p = 0.02}$) machine-learning algorithms on the estimated vGRF accuracy (RMSE). %, as well as on the correlation coefficient R, 
Similarly, the two-way ANOVA  of the RMSE obtained for the inter-participant testing showed that there was a significant effect of the algorithm type on the RMSE (F(22) = 54.31, $\mathrm{p < 0.001}$). The post hoc analysis showed that there were significant differences between the ANN and RF ($\mathrm{p = 0.025}$), ANN and LSTM ($\mathrm{p < 0.0001}$), and RF and LSTM ($\mathrm{p < 0.0001}$) machine-learning algorithms on the estimated vGRF accuracy (RMSE).
These results highlighted the superior performance of the LSTM model over the other models. Further analysis showed that CoPS (T2) has better performance than IMU (T1) data and the fusion of CoPS and IMU features (T3) resulted in the best accuracy for LSTM models. %The analysis results showed importance of data fusion in enhancing vGRF estimation accuracy. %and suggest that integrating multiple sensor data provides a more comprehensive representation of the biomechanical influencing factors influencing human gait, leading to more precise and robust vGRF estimations.

 %There were significant differences between ANN and RF ($\mathrm{p = 0.011}$), ANN and LSTM ($\mathrm{p < 0.0001}$), and RF and LSTM ($\mathrm{p < 0.0001}$) machine-learning algorithms on the estimated vGRF correlation coefficient (R). Based on the ANOVA results, the input feature type had no significant effect on the RMSE. 

%When we analyzed the effects of the input feature types on the RMSE for the inter-participant testing,
As the LSTM models with the IMU features, CoPS features, and the fusion of both the IMU and CoPS features have the best performances for the intra-participant and inter-participant test cases, these three LSTM models are further evaluated with NRMSE, correlation, and the accuracy of WAP and POP.

\begin{figure*}[tb!]
    \centering
    \includegraphics[width=\textwidth]{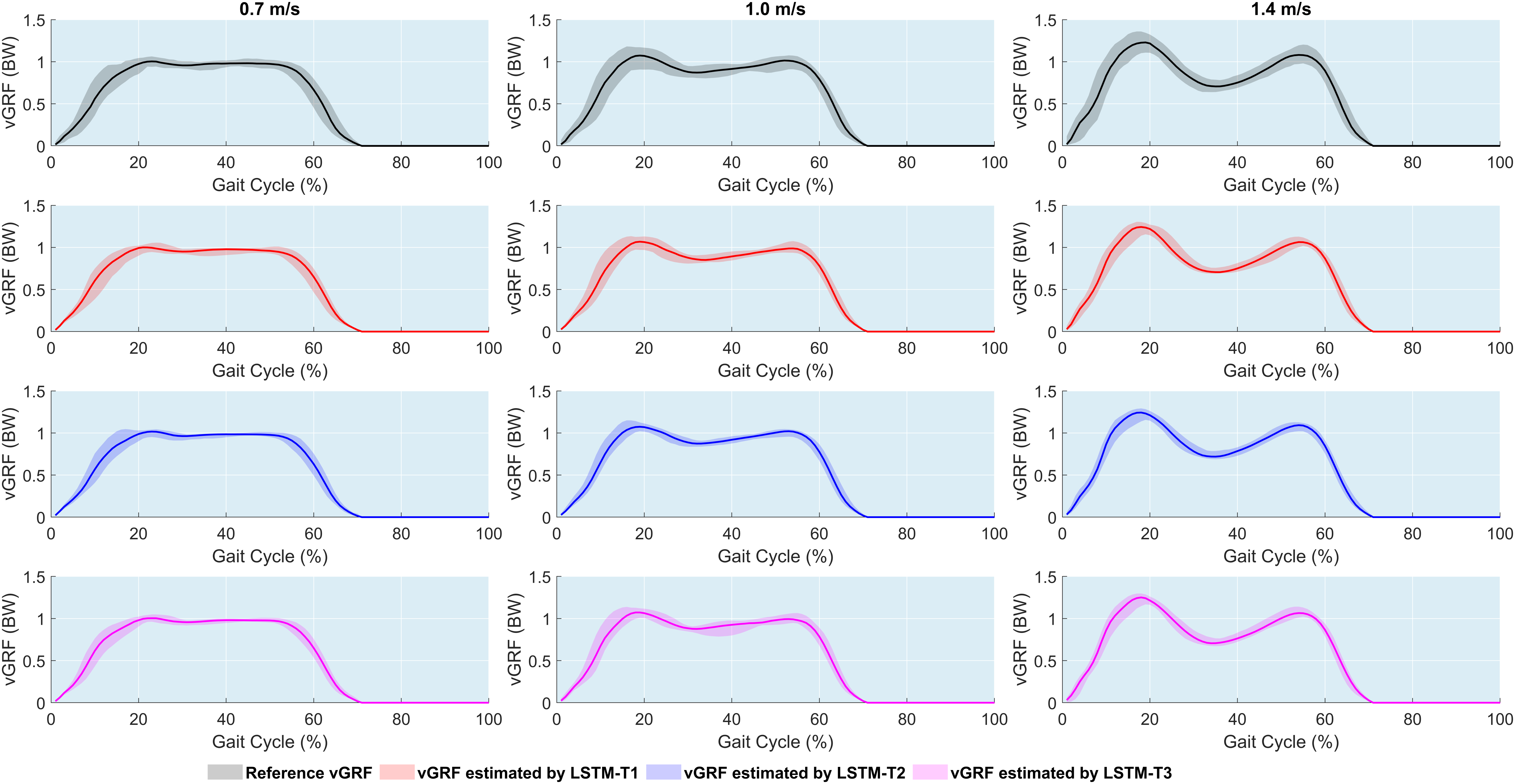}
    \caption{Epoch graphs of the measured and estimated vGRF for the inter-participant testing method at different walking speeds. The median value has been plotted with lines, while the values between the $\mathrm{2.5^{th}}$ and $\mathrm{97.5^{th}}$ percentiles are shaded. %The first row corresponds to the reference vGRF measured by the treadmill; the second-row plots are for the vGRF estimated by the LSMT-CoPS model (T2); and the third row presents the vGRF estimated using the LSTM-Fusion model (T3). The plots include all participant's data for the inter-participant testing method.
    }
    \label{fig:grf_reconstruction}
\end{figure*}

\begin{table}[htbp!]
\centering
\begin{threeparttable}
\caption{Correlation (R) and normalized RMSE (NRMSE) results between the reference vGRF and the estimated vGRF by the LSTM models for the intra-participant and inter-participant testing.}

\begin{tabularx}{\columnwidth}{|c|c|>{\centering\arraybackslash}X|>{\centering\arraybackslash}X|>{\centering\arraybackslash}X|>{\centering\arraybackslash}X|}
\hline

\multirow{2}{*}{\textbf{LSTM}} & \textbf{Speed} & \multicolumn{2}{c|}{\textbf{Intra-Participant}} & \multicolumn{2}{c|}{\textbf{Inter-Participant}}  \\ \cline{3-6}

  & \textbf{(m/s)} &  \textbf{R} & \textbf{NRMSE (\% BW)} &  \textbf{R} & \textbf{NRMSE (\% BW)} \\ \hline

\multirow{4}{*}{T1}         & 0.7    & 0.9950 & 2.66 & 0.9886 & 4.05  \\

                            & 1.0    & 0.9962 & 2.30 & 0.9885 & 3.98  \\

                            & 1.4    & 0.9959 & 2.16 & 0.9869 & 3.85  \\
                            
                            & Global & 0.9958 & 2.13 & 0.9878 & 3.64  \\ \hline

\multirow{4}{*}{T2}         & 0.7    & 0.9951 & 2.62 & 0.9902 & 3.76  \\

                            & 1.0    & 0.9968 & 2.10 & 0.9883 & 3.99  \\

                            & 1.4    & 0.9965 & 1.99 & 0.9880 & 3.64  \\
                            
                            & Global & 0.9964 & 1.96 & 0.9888 & 3.46  \\ \hline

\multirow{4}{*}{T3}         & 0.7    & 0.9966 & 2.19 & 0.9911 & 3.55  \\

                            & 1.0    & 0.9974 & 1.91 & 0.9900 & 3.67  \\

                            & 1.4    & 0.9969 & 1.89 & 0.9906 & 3.27  \\
                            
                            & Global & 0.9970 & 1.79 & 0.9905 & 3.22  \\ \hline
                            
\end{tabularx}
\begin{tablenotes}
\item Note: T1 corresponds to the LSTM model trained with the IMU feature, T2 to the LSTM model with the CoPS feature, and T3 to the LSTM model with the data fusion of IMU and CoPS features.
\end{tablenotes}
\label{tab:result_summary}
\end{threeparttable}
\end{table}

\begin{table}[htbp!]
\centering
\begin{threeparttable}
\caption{WAP and POP errors obtained with the LSTM models for the intra-participant and inter-participant testing.}

\begin{tabularx}{\columnwidth}{|c|c|>{\centering\arraybackslash}X|>{\centering\arraybackslash}X|>{\centering\arraybackslash}X|>{\centering\arraybackslash}X|}
\hline

\multirow{2}{*}{} & \multirow{2}{*}{} & \multicolumn{2}{c|}{\textbf{Intra-parcitipant}} & \multicolumn{2}{c|}{\textbf{Inter-parcitipant}} \\ \cline{3-6}

 \textbf{LSTM} & \textbf{Speed (m/s)} & \textbf{WAP (BW)} & \textbf{POP (BW)} & \textbf{WAP (BW)} & \textbf{POP (BW)}  \\  \hline

\multirow{4}{*}{T1}         & 0.7    & 0.023 & 0.017 & 0.026 & 0.021 \\

                            & 1.0    & 0.032 & 0.021 & 0.055 & 0.033 \\

                            & 1.4    & 0.032 & 0.024 & 0.059 & 0.039 \\

                            & Global & 0.030 & 0.022 & 0.051 & 0.033 \\ \hline

\multirow{4}{*}{T2}         & 0.7    & 0.017 & 0.014 & 0.021 & 0.014 \\

                            & 1.0    & 0.024 & 0.014 & 0.059 & 0.027 \\

                            & 1.4    & 0.028 & 0.021 & 0.055 & 0.035 \\

                            & Global & 0.024 & 0.017 & 0.050 & 0.028 \\ \hline

\multirow{4}{*}{T3}         & 0.7    & 0.016 & 0.012 & 0.021 & 0.019 \\

                            & 1.0    & 0.020 & 0.014 & 0.049 & 0.021 \\

                            & 1.4    & 0.026 & 0.021 & 0.051 & 0.031 \\

                            & Global & 0.022 & 0.017 & 0.044 & 0.025 \\ \hline
                            
\end{tabularx}
\begin{tablenotes}
\item Note: T1 corresponds to the LSTM model trained with the IMU feature, T2 to the LSTM model with the CoPS feature, and T3 to the LSTM model with the data fusion of IMU and CoPS features.
\end{tablenotes}
\label{tab:result_peak_error}
\end{threeparttable}
\end{table}

\begin{table}[htbp!]
\centering
\begin{threeparttable}
\caption{WAP and POP timing delay obtained with the LSTM models for the intra-participant and inter-participant testing.}

\begin{tabularx}{\columnwidth}{|c|c|>{\centering\arraybackslash}X|>{\centering\arraybackslash}X|>{\centering\arraybackslash}X|>{\centering\arraybackslash}X|}
\hline

\multirow{2}{*}{} & \multirow{2}{*}{} & \multicolumn{2}{c|}{\textbf{Intra-parcitipant}} & \multicolumn{2}{c|}{\textbf{Inter-parcitipant}} \\ \cline{3-6}

 \textbf{LSTM} & \textbf{Speed (m/s)} & \textbf{WAP (\% GC)} & \textbf{POP (\% GC)} & \textbf{WAP (\% GC)} & \textbf{POP (\% GC)}  \\  \hline

\multirow{4}{*}{T1}     & 0.7    & 1.6 & 4.0 & 3.3 & 4.4 \\

                        & 1.0    & 1.3 & 2.1 & 2.0 & 2.7 \\

                        & 1.4    & 0.9 & 0.7 & 1.3 & 1.2 \\

                        & Global & 1.2 & 1.9 & 2.0 & 2.4 \\ \hline

\multirow{4}{*}{T2}     & 0.7    & 2.7 & 3.3 & 2.5 & 4.2 \\

                        & 1.0    & 1.3 & 2.0 & 1.9 & 2.2 \\

                        & 1.4    & 0.8 & 0.7 & 1.1 & 1.0 \\

                        & Global & 1.4 & 1.7 & 1.7 & 2.1 \\ \hline

\multirow{4}{*}{T3}     & 0.7    & 2.9 & 3.9 & 2.3 & 4.3 \\

                        & 1.0    & 1.2 & 1.8 & 1.6 & 2.7 \\

                        & 1.4    & 0.8 & 0.6 & 1.0 & 0.9 \\

                        & Global & 1.4 & 1.8 & 1.5 & 2.3 \\ \hline
                            
\end{tabularx}
\begin{tablenotes}
\item Note: T1 corresponds to the LSTM model trained with the IMU feature, T2 to the LSTM model with the CoPS feature, and T3 to the LSTM model with the data fusion of IMU and CoPS features. GC -- gait cycle.
\end{tablenotes}
\label{tab:result_peak_delay}
\end{threeparttable}
\end{table}

%\subsection{Performance of LSTM Models for vGRF Estimation} \label{lstm_model_performance}
Figure \ref{fig:grf_reconstruction} shows the reference vGRF, the vGRF estimated with the LSTM-T1, LSTM-T2, and LSTM-T3. The solid continuous line indicates the median value of the vGRF for all gait cycles and the shaded region corresponds to the values between $\mathrm{2.5^{th}}$ and $\mathrm{97.5^{th}}$ percentiles of the vGRF signals. Table \ref{tab:result_summary} shows the correlation coefficient (R) and NRMSE between the reference and estimated vGRF with the selected models (LSTM-T1, LSTM-T2, and LSTM-T3) for the intra-participant and inter-participant testing. Table \ref{tab:result_peak_error} and Table \ref{tab:result_peak_delay} show the mean magnitude errors and delays of the estimated vGRF peaks (WAP and POP). Overall, the LSTM-T2 has a better performance than LSTM-T1 which shows the advantage of the CoPS over IMU in estimating vGRF. For most cases, LSTM-T3 has the best performance which shows the advantage of fusing CoPS and IMU data in estimating vGRF. 

%The results for the intra-participant testing for the three LSTM models showed R values for the estimated vGRF between 0.988 and 0.9974, with RMSE and NRMSE ranging between 0.023 BW to 0.052 BW and 1.89\% to 3.99\%, respectively (Table \ref{tab:result_summary}). The analysis of the RMSE in the estimated vGRF peaks for the LSTM models ranges from 0.016 BW to 0.059 BW for the WAP and between 0.012 BW and 0.039 BW for the POP peaks. Similar accuracies are observed for the timing delays in the vGRF peak estimation (WAP -- 0.8\% to 3.3\%, and POP -- 0.6\% to 4.4\%) for intra- and inter-participant testings. %Post hoc statistical analysis showed no significant differences between the performance of the three LSTM models across all walking gait speeds.

% ******************************************************* Section End

\section{Discussion} \label{discussion}
This study introduced new data -- CoPS for vGRF estimating and showed its advantage over existing IMU-based methods. Additionally, other than popularly used ANN and RF, this study also evaluated LSTM and showed its advantage in vGRF estimation. 

%Additionally, we investigated the performance of three machine learning algorithms (ANN, RF, and LSTM) to estimate vGRF. The CoPS estimation was based on an insole pressure-sensing array operated as foot contact switches. 
%As CoPS has One of the main advantages of the CoPS estimation method is that the estimated center of the pressed sensor does not rely on continuous pressure values. Therefore, sensor drift and hysteresis associated with insole pressure-sensing arrays do not influence the error made in the estimation of vGRF.

%The machine-learning models were trained, and their effectiveness in estimating vGRF was evaluated during three gait speeds for intra-participant and inter-participant tests. In both testing scenarios, the LSMT model outperformed the other machine-learning models. The highest accuracies were achieved for the LSTM-T3 model, with correlation coefficients (R) of 0.9974 and 0.9911 between the estimated vGRF and the reference measurements in the intra-participant and inter-participant tests, respectively. The results demonstrated the feasibility of the proposed method for vGRF estimation during walking at different speeds.

Through comparing the performances of all nine trained models (ANN, RF, and LSTM models for T1, T2, and T3 features), some differences can be observed between them for the intra- and inter-participant testing. The RMSE of the ANN model ranges between 0.049 BW and 0.063 BW, which is about twice the RMSE of RF and LSTM models (0.023 BW to 0.041 BW) during the intra-participant testing. This can be attributed to the inability of the ANN model to extract the temporal dependency in the walking gait data and a similar problem was reported in \cite{martinez2023estimating}. Despite the comparable performance of the RF model with the LSTM models for the intra-participant testing, there seems to be a significant performance degradation during inter-participant testing (Table \ref{tab:result_intra} and Table \ref{tab:result_inter}). The performance degradation of the RF models can be attributed to model overfitting during training, a common problem with algorithms based on decision trees \cite{dietterich1995overfitting}. %Despite the overfitting associated with the RF model, the result showed comparable performances (global RMSE, 0.070 BW and correlation (R), 0.0624) with those reported in previous studies \cite{jiang2020estimating, martinez2023estimating}. 
%Further analysis of the LSTM-T1, LSTM-T2, and LSTM-T3 models shows that the reference vGRF and the estimated vGRF from the three models are strongly correlated for both the intra- and inter-participant testings (R ranges between 0.988 and 0.9974). Similar results can be seen in the results obtained for the RMSE (0.023 BW to 0.052 BW) and NRMSE(1.91\% to 4.05\%) across all gait speeds, respectively.

The ANOVA analysis for the intra-participant and inter-participant testing reveals that the LSTM model consistently outperforms the ANN and RF models, as evidenced by lower RMSE, NRMSE, and higher correlation coefficients. This performance can be attributed to the bi-directional LSTM's architecture, which effectively captures temporal dependencies inherent in gait data in forward and reverse directions. The superior accuracy and reduced error rates suggest that the LSTM model is particularly well-suited for applications requiring precise vGRF estimation. %Furthermore, integrating CoPS and IMU data in the LSTM-T3 model enhanced prediction accuracy by providing complementary information about foot pressure distribution and movement dynamics. Future work should investigate the individual contributions of these features in greater detail, potentially leading to further model refinement.

\begin{table}[htbp!]
\centering
\begin{threeparttable}
\caption{Inter-Participant Performance Comparison of vGRF Estimation With Existing Methods.}

\begin{tabularx}{\columnwidth}{|l|c|>{\centering\arraybackslash}X|>{\centering\arraybackslash}X|>{\centering\arraybackslash}X|>{\centering\arraybackslash}X|}
\hline

\textbf{Author} & \textbf{Year} & \textbf{Model}& \textbf{RMSE (BW)} & \textbf{NRMSE (\%BW)}  \\ \hline

Our method                                                 &   2024   &  LSTM   &  0.044  &  3.22 \\ \hline

Jiang et al. \cite{jiang2020estimating}                    &  2020  &  RF     &  0.100  &  7.15  \\ \hline

Martinez-Pascual et al. \cite{martinez2023estimating}      &  2023  &  ANN    &  0.074  &  6.84  \\ \hline

Martinez-Pascual et al. \cite{martinez2023estimating}      &  2023  &  RF     &  0.077  &  7.27  \\ \hline

Karatsidis et al. \cite{karatsidis2016estimation}   &  2017  &  HBM/NE  &  0.063  &  5.30  \\ \hline
                            
\end{tabularx}
\begin{tablenotes}
\item Note: The unit of RMSE is body weight (BW), NRMSE -- percent of the body weight (\%BW). HBM -- Human Biomechanical model, and NE -- Newton-Euler Equation.
\end{tablenotes}
\label{tab:result_comparison}
\end{threeparttable}
\end{table}

The proposed vGRF estimation methods proposed in this study outperformed the existing state-of-the-art solutions. Table \ref{tab:result_comparison} shows the comparison between our proposed method and existing methods reported in the literature. Martinez-Pascual et al. \cite{martinez2023estimating} investigated using an ANN and RF to estimate vGRF with five IMU sensors during walking. The best RMSE and NRMSE achieved were 0.074 BW and 6.84\% BW, respectively. Jiang et al. \cite{jiang2020estimating} used RF to achieve RMSE and NRMSE of 0.1 BW and 7.15\% BW, respectively. Karatsidis et al. \cite{karatsidis2016estimation} used human biomechanical models and Newton-Euler's methods to achieve 0.063 BW and 5.3\% BW for RMSE and NRMSE, respectively.  % Other studies have used decision tree regression approaches to estimate the vGRF during walking with NRMSE reported between 6.84\% and 7.15\% \cite{jiang2020estimating,martinez2023estimating} for inter-participant testing. 
The method proposed in this study, which estimates vGRF with LSTM and CoPS and IMU data, achieved 0.044 BW and 3.22\% BW for RMSE and NRMSE, respectively. This result outperformed vGRF estimation methods previously reported in the literature \cite{ancillao2018indirect,shahabpoor2017measurement}. 

%The performance of machine-learning models on unseen user data during training is crucial for enhancing system practicality, allowing for cross-participant assessments without the need to train a new model for each user. A wearable device utilizing the proposed method should come equipped with a pre-trained regression model before reaching the end user. This is vital in real-world scenarios where users typically lack the skills and facilities to generate the reference data necessary for training a personalized model. In this study's inter-participant testing context, we employed a leave-one-subject-out cross-validation technique to assess model generalization. This approach ensures that data excluded from training is used for testing, demonstrating the robustness and applicability of the pre-trained model across all users without the need for retraining.

This study was limited to the vGRF estimation for only healthy adult populations. Variations in age, foot shape, and gait pathologies might introduce complexities that the current method may not fully capture. Future works will explore a wider population demographic, including populations with a wider range of ages, heights, sexes, and distinct gait patterns, such as elderly individuals or those with neurological conditions.

% ******************************************************* Section End

\section{Conclusion} \label{conclusion}
This paper presents a method of estimating the vGRF with CoPS and IMU. Three machine learning models: ANN, RF, and LSTM were evaluated for vGRF estimation. The results showed that CoPS has better performance than IMU and LSTM has better performance over popularly used ANN and RF. The LSTM model trained with the fusion of CoPS and IMU data achieved the best performance when compared with existing state-of-the-art solutions. The results of this study show the feasibility of using the CoPS and IMU for accurate vGRF estimation during walking. %proposed method achieves a very high correlation between the estimated vGRF and the reference measurement (R = 0.9974) and an RMSE of 0.024 BW, average all gait speeds. The timing of the weight acceptance peaks were identified accurately within 0.8\% and 2.9\%, and push-off peaks within 0.7\% and 3.9\% of the gait cycle duration in both the intra-participant and inter-participant testing, respectively. 

% ******************************************************* Section End

\section*{Acknowledgments}
We would like to express our gratitude to Professor Dain LaRoche and the Kinesiology Department of the University of New Hampshire for providing access to the instrumented treadmill, which was essential for conducting experiments and collecting ground truth vGRF data. %This should be a simple paragraph before the References to thank those individuals and institutions who have supported your work on this article.

% \begin{thebibliography}
\bibliography{reference}
\bibliographystyle{IEEEtran}

\vfill

\end{document}